\pdfoutput=1

\documentclass[11pt]{article}

\usepackage[preprint]{acl}

\usepackage{times}
\usepackage{latexsym}

\usepackage[T1]{fontenc}

\usepackage[utf8]{inputenc}

\usepackage{microtype}

\usepackage{inconsolata}

\usepackage{graphicx}

\usepackage{epstopdf}
\usepackage{subfigure}
\usepackage{amsthm}
\usepackage{enumitem}
\usepackage[normalem]{ulem}
\usepackage{algorithm}
\usepackage{algpseudocode}
\usepackage[most]{tcolorbox}
\usepackage{xcolor} 
\usepackage{colortbl}
\usepackage{bbm}
\usepackage{bm}
\useunder{\uline}{\ul}{}
\graphicspath{{./fig/}}

\usepackage{balance}
\usepackage{subcaption}
\usepackage{amsmath}
\usepackage{tabularray}

\usepackage{array}          
\usepackage{booktabs}       
\usepackage{arydshln}       
\usepackage{longtable}      
\usepackage{multirow}       
\title{Inference Computation Scaling for Feature Augmentation in Recommendation Systems}

\author{
  Weihao Liu\textsuperscript{1}, 
  Zhaocheng Du\textsuperscript{2},
  Haiyuan Zhao\textsuperscript{1}, 
  Wenbo Zhang\textsuperscript{1}, 
  Xiaoyan Zhao\textsuperscript{3}, \\
  \textbf{Gang Wang\textsuperscript{2},
  Zhenhua Dong\textsuperscript{2},
  Jun Xu\textsuperscript{1}\thanks{Corresponding author.} }\\
  \textsuperscript{1}Renmin University of China \\
  \textsuperscript{2}Huawei Noah's Ark Lab\\
  \textsuperscript{3}The Chinese University of Hong Kong \\
  \texttt{\{weihaoliu, haiyuanzhao, junxu\}@ruc.edu.cn, xzhao@se.cuhk.edu.hk} \\
    \texttt{\{zhaochengdu, wanggang110, dongzhenhua\}@huawei.com} \\
}

\begin{document}
\maketitle
\begin{abstract}

Large language models have become a powerful method for feature augmentation in recommendation systems. However, existing approaches relying on quick inference often suffer from \emph{incomplete feature coverage} and \emph{insufficient specificity} in feature descriptions, limiting their ability to capture fine-grained user preferences and undermining overall performance. 
Motivated by the recent success of \emph{inference scaling} in math and coding tasks, we explore whether scaling inference can address these limitations and enhance feature quality.

Our experiments show that scaling inference leads to significant improvements in recommendation performance, with a 12\% increase in NDCG@10. The gains can be attributed to two key factors: feature quantity and specificity. In particular, models using extended Chain-of-Thought (CoT) reasoning generate a greater number of detailed and precise features, offering deeper insights into user preferences and overcoming the limitations of quick inference. We further investigate the factors influencing feature quantity, revealing that model choice and search strategy play critical roles in generating a richer and more diverse feature set. This is the first work to apply inference scaling to feature augmentation in recommendation systems, bridging advances in reasoning tasks to enhance personalized recommendation.

\end{abstract}

\section{Introduction}

\begin{figure}[t]
  \includegraphics[width=\columnwidth]{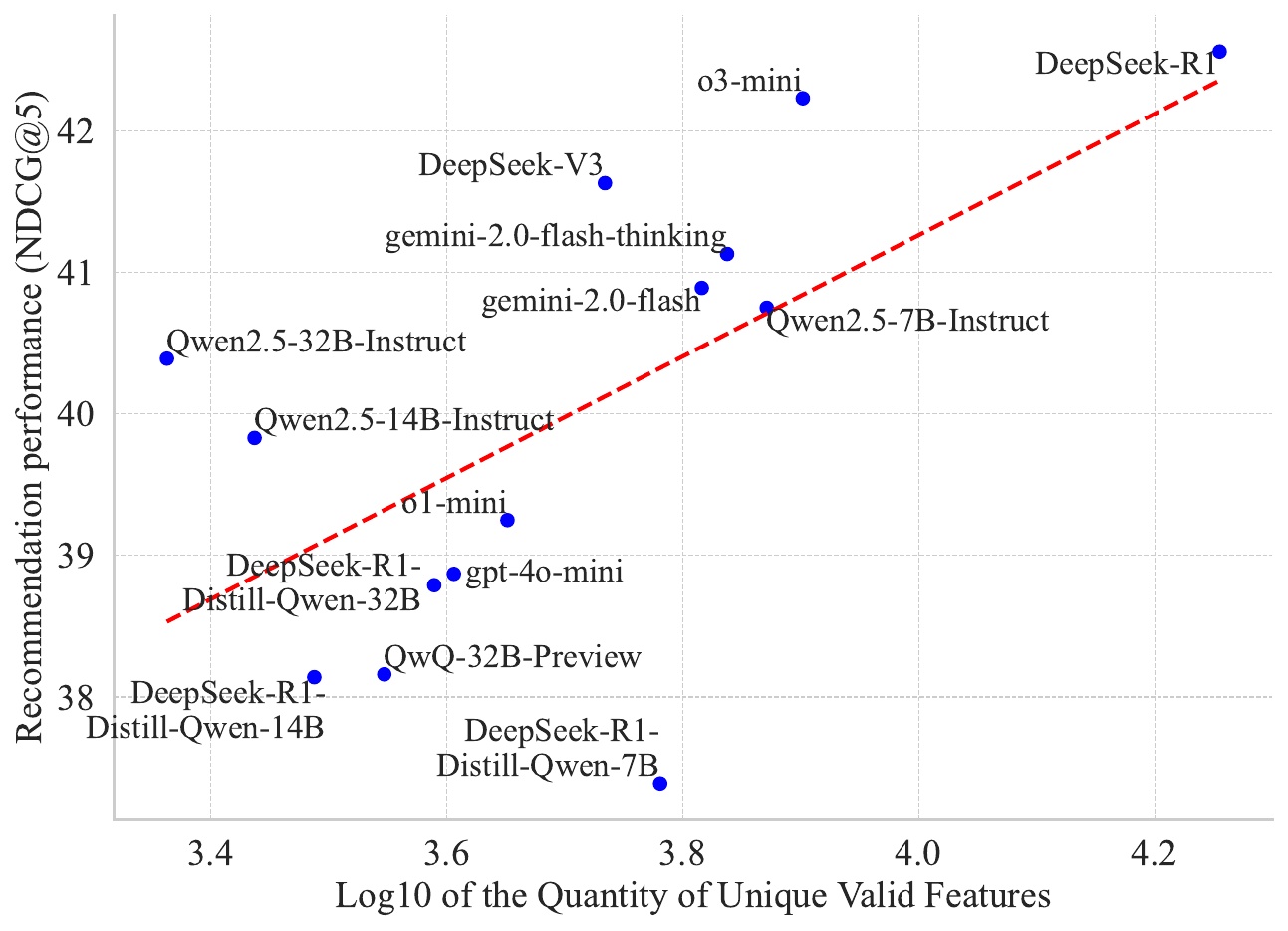}
  \caption{The positive correlation between recommendation performance and the number of unique valid features generated by different LLMs. The red dotted line represents the best-fit line of the data.}
  \label{fig:performance_wrt_count}
\end{figure}

Large language models (LLMs) have recently become a prevalent and effective approach to feature augmentation in recommendation systems~\citep{wu2024survey, wei2024llmrec}. By leveraging their generative capabilities, LLMs can produce a range of item~\citep{acharya2023llm, ren2024representation} or user-level~\citep{wang2023zero, wang2023enhancing} features—ranging from high-level product attributes~\citep{xi2024towards} to nuanced decision factors~\citep{sun2024large}—thereby enhancing recommendation performance. However, most existing methods rely on quick inference (also called System-1 thinking)~\citep{kahneman2011thinking} to generate these features, which often leads to issues such as incomplete feature coverage and insufficient specificity~\citep{ji2025test}. As a result, these fast-thinking models lack depth analysis and some critical dimensions of user preferences remain unexplored, so the resulting features may fall short in capturing the subtle decision-making factors that differentiate preferred items from disliked ones.

A concrete example illustrates this limitation: when using a standard LLM, like gpt-4o-mini~\citep{gpt_4o_mini}, to analyze a user's interaction history with musical instruments, the model generates a vague feature such as ``Component Functionality.'' However, this feature lacks the specific details that truly matter to a musician’s decision-making process, like ``Enhanced Durability,'' ``Surface Look,'' and ``Material Type.'' In contrast,  
by employing extended Chain-of-Thought (CoT) reasoning~\citep{wei2022chain} with models like o1-mini~\citep{o1_system_card} or o3-mini~\citep{o3}, these more detailed and personalized features are uncovered, resulting in a richer and more precise feature set.  
This example highlights how deeper reasoning allows LLMs to better capture the nuances of user preferences, emphasizing the need for inference scaling to overcome limitations of quick thinking.

Inspired by the remarkable success of \emph{inference computation scaling}~\citep{snell2024Scaling, o1_blog} in tasks such as math~\citep{wang2024openr} and coding~\citep{o1-coder}—where allocating more computational resources yields better performance~\citep{zeng2024scaling}—this work investigates whether these same principles can be harnessed for feature augmentation in recommendation. Our goal is to determine whether extended reasoning chains can discover more diverse and interpretable user-level features, thereby improving downstream recommendation performance. Concretely, we address the following three research questions (RQs):

\textbf{RQ1: Can inference scaling improve recommendation performance, and if so, to what extent?}
We compare baseline LLMs (e.g., gpt-4o-mini) with models that use extended reasoning (e.g., o1-mini and o3-mini). Our experiments show that models with long-CoT reasoning outperform those without, achieving a 12\% increase in NDCG@10. 
This demonstrates that extended reasoning, which has proven effective in math and coding tasks, can also enhance recommendation models.
Interestingly, while algorithms like Beam Search~\citep{xie2023self} and Monte Carlo Tree Search (MCTS)~\citep{browne2012survey} have achieved significant success in other domains, they do not perform as well as the Best-of-N~\citep{cobbe2021training} strategy for feature augmentation in recommendation tasks. These findings suggest that deeper inference, combined with the appropriate search strategy, can lead to substantial improvements in recommendation performance.

\textbf{RQ2: Which factors drive these performance gains?}
We observe that the improvements in recommendation performance are largely driven by the increased feature quantity and enhanced specificity achieved through inference scaling. In our study, feature quantity serves as a proxy for reasoning depth—more unique valid features indicate a deeper exploration of user preferences, thereby addressing the issue of incomplete feature coverage. To assess this, we plot the number of unique valid features generated by different LLMs against their recommendation performance, as shown in Figure~\ref{fig:performance_wrt_count}. Our results reveal a positive correlation: models like DeepSeek-R1~\citep{deepseek_r1} and o3-mini, which generate the most features, also deliver the best performance. In addition, using an LLM-as-a-judge~\citep{zheng2024judging} approach, we find that inference scaling models consistently produce more detailed and precise features compared to non-scaled models, effectively mitigating the problem of insufficient specificity.

\textbf{RQ3: Given that feature quantity influences performance, what factors affect the number of generated features?}
We investigate how model families, long-CoT reasoning, and model size impact the overall quantity of generated features. Additionally, we explore the effect of search strategies. Our results show that feature generation is not merely the sum of individual steps—pursuing local optimality at the step level does not always lead to overall optimality. Consequently, step-level methods like Beam Search and MCTS underperform compared to the solution-level Best-of-N approach in producing a larger number of useful features. These findings highlight the important interaction between models and search strategies in uncovering a more comprehensive set of features.

In summary, our contributions are as follows:

(1) To the best of our knowledge, we are the first to explore inference scaling to address limitations in feature augmentation for recommendation systems, while also evaluating performance improvements across various LLMs.

(2) We investigate how feature quantity and specificity explain the performance gains achieved through inference scaling.

(3) We provide an in-depth analysis of how search algorithms influence feature generation, highlighting their role in enhancing performance.

\section{Preliminaries}


\subsection{Task Formulation}

We adopt a typical sequential recommendation setting~\citep{fang2020deep}, where each user's interactions are sorted in chronological order. For each user \(u\), the final interacted item is set aside as the test item, while all previously interacted items constitute the training set~\citep{he2017neural}. The goal is to analyze the training interactions—inferring the factors that differentiate items the user likes from those the user does not—and then predict the held-out test item (i.e., the next item they will interact with) from a set of candidate items.

\subsection{Experimental Setup}

\subsubsection{Datasets} 
We conducted experiments on two public datasets: \textbf{Toys} and \textbf{Instruments}, representing the ``Toys and Games'', and ``Musical Instruments'' categories from the Amazon dataset~\citep{ni2019justifying}, respectively. The characteristics of them are summarized in Appendix Table~\ref{tab:stats_datasets}. We use the dataset in a manner consistent with its original intended use.


\subsubsection{Key Components of Inference Scaling} 

We frame our inference scaling procedure from a reinforcement learning (RL) perspective~\citep{zeng2024scaling} with the following elements:

\textbf{Policy Model} is responsible for generating features that capture user preferences. 
   With the prompt shown in Appendix Figure~\ref{fig:PromptTemplate}, the policy model is prompted to analyze the differences between items the user likes and dislikes and generate potential features for recommendation systems. 

\textbf{Reward model} evaluates all features generated by the policy model, designating them as \emph{valid} if they successfully distinguish between a user’s liked and disliked items. Only those valid features are used to enhance downstream recommendation performance.
Specifically, we use Qwen2.5-7B-Instruct~\citep{qwen_2.5} as the reward model. While one could, in principle, use downstream recommendation performance as a reward signal, doing so would require an extensive and costly process of item-level feature completion for each candidate feature. As the focus of this paper is on the use of inference computation for feature augmentation, we leave the more complex tasks of designing reward functions and fine-tuning LLMs based on rewards for future research.

\textbf{Search Strategy} guides how the policy model explores and selects the generated features. Various strategies can be employed, such as Best-of-N~\citep{cobbe2021training}, Beam Search~\citep{xie2023self}, or Monte Carlo Tree Search (MCTS)~\citep{browne2012survey}. These strategies strike a balance between exploring new features and exploiting features that have already shown promise in previous steps. By carefully selecting features with high rewards, the model can efficiently navigate the vast space of potential features, ensuring that the most effective features are identified and used for the final recommendation task.

\section{Impact of Inference Scaling on Recommendation Performance (RQ1)}

\begin{table*}
\caption{Performance of features from various policy models on downstream recommendation tasks. The baseline performance, labeled as ``w/o features'', reflects results without the inclusion of additional features. Bold denotes the highest scores across all policy models. For brevity, `\%' is omitted from scores in subsequent tables and figures.}
\label{tab:main_res}
\centering
\resizebox{\textwidth}{!}{%
\begin{tabular}{c|l|ccccc|ccccc|c}
\hline\hline
\multirow[m]{2}{*}{Dataset}      & \multirow{2}{*}{Policy Model} & \multicolumn{5}{c|}{DR}                        & \multicolumn{5}{c|}{ICL}                        & NIP \\ \cline{3-13} 
    &     & Valid Rate & NDCG@5 & HIT@5 & NDCG@10 & HIT@10 & Valid Rate & NDCG@5 & HIT@5 & NDCG@10 & HIT@10 & HIT     \\ 
\hline
\multirow[m]{15}{*}{Toys}        & w/o features                  & 94.80            & 37.21  & 51.01 & 43.50    & 70.67  & 99.33           & 35.46  & 49.69 & 42.06   & 70.28  & 24.46   \\
    & Qwen2.5-7B-Instruct           & 97.28           & 40.75  & 56.72 & 46.84   & 75.54  & 99.67           & 39.26  & 54.88 & 45.73   & 74.98  & 27.13   \\
    & Qwen2.5-14B-Instruct          & 97.38           & 39.83  & 55.24 & 46.20    & 74.98  & 99.52           & 38.22  & 53.47 & 44.90    & 74.22  & 26.23   \\
    & Qwen2.5-32B-Instruct          & 96.95           & 40.39  & 54.94 & 46.95   & 75.41  & 99.71           & 39.46  & 55.14 & 45.91   & 75.23  & 26.75   \\
    & DeepSeek-R1-7B   & 96.90            & 37.39  & 52.36 & 44.15   & 73.33  & 99.52           & 36.24  & 51.37 & 43.09   & 72.69  & 24.27   \\
    & DeepSeek-R1-14B  & 95.90            & 38.14  & 52.56 & 44.71   & 73.05  & 99.57           & 35.81  & 49.86 & 43.12   & 72.65  & 25.08   \\
    & DeepSeek-R1-32B  & 97.38           & 38.79  & 53.97 & 45.58   & 75.02  & 100.00            & 37.11  & 52.84 & 43.85   & 73.63  & 25.56   \\
    & QwQ-32B-Preview               & 96.09           & 38.16  & 53.10  & 44.37   & 72.41  & 99.71           & 35.32  & 50.65 & 42.22   & 72.12  & 24.75   \\
    & gpt-4o-mini                        & 97.14           & 38.87  & 52.97 & 45.64   & 74.13  & 99.71           & 38.16  & 52.89 & 44.78   & 73.46  & 26.51   \\
    & o1-mini                       & 97.19           & 39.25  & 53.29 & 46.13   & 74.68  & 99.81           & 38.82  & 53.94 & 45.44   & 74.58  & 25.99   \\
    & o3-mini                       & 97.23           & 42.23  & \textbf{56.89} & 48.11   & 75.28  & 99.86           & 40.69  & 55.97 & 47.25   & \textbf{76.41}  & 29.04   \\
    & DeepSeek-V3                   & 97.28           & 41.63  & 56.47 & 48.15   & \textbf{76.72}  & 99.90            & 40.98  & 55.80  & 47.06   & 74.65  & 28.42   \\
    & DeepSeek-R1                   & 97.23           & \textbf{42.56}  & 56.74 & \textbf{48.86}   & 76.31  & 99.62           & \textbf{42.35}  & \textbf{58.83} & \textbf{47.66}   & 75.30   & \textbf{29.33}   \\
    & gemini-2.0-flash              & 97.38           & 40.89  & 55.63 & 47.12   & 75.07  & 99.76           & 40.76  & 56.64 & 46.82   & 75.48  & 27.61   \\
    & gemini-2.0-flash-thinking-exp & 97.42           & 41.13  & 55.07 & 47.35   & 74.40   & 99.76           & 41.24  & 57.31 & 47.20    & 75.72  & 28.85   \\ 
\hline
\multirow[m]{15}{*}{Instruments} & w/o features                  & 95.33           & 40.77  & 56.13 & 46.85   & 75.23  & 99.75           & 36.01  & 53.14 & 43.23   & 75.59  & 23.99   \\
    & Qwen2.5-7B-Instruct           & 97.42           & 44.10   & 58.33 & 50.24   & 77.53  & 99.63           & 39.73  & 54.44 & 46.66   & 75.80   & 27.68   \\
    & Qwen2.5-14B-Instruct          & 98.28           & 39.28  & 53.44 & 46.67   & 76.47  & 99.63           & 35.87  & 52.10  & 44.07   & 77.53  & 27.68   \\
    & Qwen2.5-32B-Instruct          & 97.79           & 43.28  & 58.11 & 49.92   & 78.62  & 99.88           & 39.95  & 57.27 & 47.13   & 79.56  & 27.68   \\
    & DeepSeek-R1-7B   & 96.56           & 41.88  & 57.96 & 48.37   & 77.96  & 99.88           & 36.80   & 53.94 & 44.45   & 77.83  & 25.22   \\
    & DeepSeek-R1-14B  & 96.43           & 40.31  & 55.23 & 47.75   & 78.44  & 99.63           & 35.55  & 52.72 & 43.01   & 76.05  & 24.48   \\
    & DeepSeek-R1-32B  & 97.66           & 40.13  & 55.04 & 46.99   & 76.20   & 99.63           & 38.66  & 55.93 & 45.23   & 76.30   & 25.46   \\
    & QwQ-32B-Preview               & 95.69           & 43.20   & 59.25 & 49.50    & 78.66  & 99.63           & 39.22  & 55.80  & 46.64   & 78.77  & 25.22   \\
    & gpt-4o-mini                        & 97.79           & 41.86  & 57.48 & 48.76   & 78.99  & 100.00            & 40.79  & 58.43 & 47.34   & 78.84  & 29.77   \\
    & o1-mini                       & 98.15           & 46.21  & 62.66 & 51.65   & 79.57  & 99.88           & 43.04  & 61.08 & 48.81   & 79.06  & 30.01   \\
    & o3-mini                       & 97.54           & 45.60   & 59.90  & 52.12   & 80.08  & 99.88           & 43.20   & 59.85 & 49.31   & 78.82  & 30.01   \\
    & DeepSeek-V3                   & 97.91           & 48.10   & 62.69 & 53.84   & 80.65  & 99.63           & 44.59  & 62.10  & 51.04   & 82.10   & 32.10    \\
    & DeepSeek-R1                   & 98.52           & 47.82  & 62.05 & 53.36   & 79.28  & 99.63           & 44.29  & 61.73 & 50.60    & 81.48  & 30.75   \\
    & gemini-2.0-flash              & 97.54           & 48.98  & 64.56 & 55.05   & \textbf{83.23}  & 99.88           & 45.65  & 64.53 & 51.54   & 82.76  & \textbf{31.49}   \\
    & gemini-2.0-flash-thinking-exp & 98.77           & \textbf{49.96}  & \textbf{65.63} & \textbf{55.50}    & 82.57  & 100.00            & \textbf{47.41}  & \textbf{64.82} & \textbf{53.54}   & \textbf{83.76}  & 31.24   \\ 
\hline\hline
\end{tabular}%
}
\end{table*}

In this section, we first investigate whether inference scaling can enhance the performance of recommendation systems. 

\subsection{Experimental Setup}

\textbf{Models.} To investigate how features from different models impact performance, we employ various LLMs as policy models. Specifically, we use the \textbf{GPT} series, including gpt-4o-mini~\citep{gpt_4o_mini}, o1-mini~\citep{o1_system_card}, o3-mini~\citep{o3}. The \textbf{Qwen} series consists of Qwen2.5-Instruct at 7B, 14B, and 32B scales~\citep{qwen_2.5}, while the \textbf{Gemini} series includes gemini-2.0-flash~\citep{gemini_2.0_flash} and gemini-2.0-flash-thinking-exp~\citep{gemini_2.0_flash}. For the \textbf{DeepSeek} series, we incorporate DeepSeek-V3~\citep{deepseek_v3} and DeepSeek-R1~\citep{deepseek_r1}. In addition, DeepSeek-R1 provides distilled variants based on Qwen in 7B, 14B, and 32B configurations, which we include in our experiments \footnote{For the sake of simplicity, DeepSeek-R1-Distill-Qwen-7B will be referred to as DeepSeek-R1-7B subsequently, and the same applies to 14B and 32B.}. 

\textbf{Recommenders.} We evaluate the effect of generated features using three different recommendation models: (1) \textbf{Direct Recommendation (DR)}~\citep{liu2023chatgpt}: A listwise ranking method where the model ranks $C$ candidate items based on the user's interaction history. (2) \textbf{In-Context Learning (ICL)}~\citep{dai2023uncovering}: Similar to DR, this approach also ranks $C$ candidate items based on the user's history, but uses a one-shot setting with an example prompt. (3) \textbf{Next Item Prediction (NIP)}~\citep{geng2022recommendation}: The model predicts the next item a user is likely to interact with, based on their history and a set of $C$ candidate items.

\textbf{Metrics.} For both DR and ICL, we report HIT@K and NDCG@K (with $K \in \{5, 10\}$) as the evaluation metrics. NDCG places higher weight on items ranked at the top, while HIT treats all positions equally and essentially measures recall. For NIP, we evaluate HIT to measure prediction accuracy. Additionally, to ensure that the LLM does not produce invalid outputs (such as missing or duplicate items), we compute the Valid Rate, which measures the proportion of compliant rankings.

\textbf{Details.} For each set of candidate items, the user’s most recent interaction is treated as the ground truth (positive sample), while \(C-1\) other items are randomly sampled as negative samples. To ensure fairness, we set \(C=20\) and use gpt-4o-mini to perform the recommendation tasks. The experiment is repeated three times, and the average value is taken.

\subsection{Effect of Different Policy Model Features}
\label{sec: RQ1_policy}

\begin{table*}
\caption{Performance of features generated by various search strategies on recommendation tasks. Bold denotes the highest scores across all strategies.}
\label{tab:search_res}
\centering
\resizebox{\textwidth}{!}{%
\begin{tabular}{ll|ccccc|ccccc|c}
\hline\hline
\multicolumn{1}{l|}{\multirow{2}{*}{\begin{tabular}[c]{@{}l@{}}Policy \\ Model\end{tabular}}}         & \multirow{2}{*}{\begin{tabular}[c]{@{}l@{}}Search \\ Strategy\end{tabular}} & \multicolumn{5}{c|}{DR}                                                        & \multicolumn{5}{c|}{ICL}                                                       & NIP            \\ \cline{3-13} 
\multicolumn{1}{l|}{}                    &                & Valid Rate & NDCG@5         & HIT@5          & NDCG@10        & HIT@10         & Valid Rate & NDCG@5         & HIT@5          & NDCG@10        & HIT@10         & HIT            \\ \hline
\multicolumn{2}{c|}{w/o features}                         & 95.33      & 40.77          & 56.13          & 46.85          & 75.23          & 99.75      & 36.01          & 53.14          & 43.23          & 75.59          & 23.99          \\ \hline
\multicolumn{1}{l|}{\multirow{4}{*}{\begin{tabular}[c]{@{}l@{}}Qwen2.5-7B\\ -Instruct\end{tabular}}}  & CoT            & 97.42      & 44.10           & 58.33          & 50.24          & 77.53          & 99.63      & 39.73          & 54.44          & 46.66          & 75.80           & 27.68          \\
\multicolumn{1}{l|}{}                    & Best-of-N      & 97.54      & \textbf{44.91} & \textbf{60.66} & \textbf{51.22} & \textbf{80.20}  & 99.75      & 39.70           & 56.23          & 46.64          & 77.68          & \textbf{28.04} \\
\multicolumn{1}{l|}{}                    & Beam Search    & 97.42      & 43.70           & 59.09          & 49.85          & 78.41          & 99.63      & \textbf{40.06} & \textbf{56.67} & \textbf{46.92} & \textbf{77.90}  & 26.20           \\
\multicolumn{1}{l|}{}                    & MCTS           & 97.42      & 41.87          & 56.44          & 48.12          & 75.88          & 99.51      & 39.43          & 56.37          & 45.81          & 76.27          & 26.57          \\ \hline
\multicolumn{1}{l|}{\multirow{4}{*}{\begin{tabular}[c]{@{}l@{}}Qwen2.5-14B\\ -Instruct\end{tabular}}} & CoT            & 98.28      & 39.28          & 53.44          & 46.67          & 76.47          & 99.63      & 35.87          & 52.10           & 44.07          & \textbf{77.53} & \textbf{27.68} \\
\multicolumn{1}{l|}{}                    & Best-of-N      & 97.91      & \textbf{43.11} & \textbf{58.54} & \textbf{49.22} & 77.76          & 99.88      & \textbf{40.06} & \textbf{56.77} & \textbf{46.73} & 77.34          & \textbf{27.68} \\
\multicolumn{1}{l|}{}                    & Beam Search    & 97.79      & 41.09          & 55.97          & 47.79          & 76.73          & 99.75      & 36.62          & 52.90           & 43.93          & 75.59          & 27.31          \\
\multicolumn{1}{l|}{}                    & MCTS           & 97.79      & 40.64          & 54.72          & 48.17          & \textbf{77.99} & 99.51      & 37.99          & 54.39          & 44.95          & 76.14          & 24.72          \\ 
\hline\hline
\end{tabular}%
}
\end{table*}

We compare the effects of features generated by different policy models on recommendation performance across two datasets, as shown in Table~\ref{tab:main_res}. Key findings include:

\textbf{(1) Positive Impact of Features:}
   Overall, adding distinguishable features that capture user preferences improves the recommendation performance of the LLM. For instance,  DeepSeek-R1 achieves an NDCG@10 of 48.86 in DR, which is a 12\% improvement over the baseline (43.50). Furthermore, the Valid Rate improves after feature augmentation, indicating that the features validated by the reward model are helping the LLM complete the recommendation task more effectively.

\textbf{(2) Long-CoT vs. Non-Long-CoT Models:}
   In our experiments, we observe that models with long-CoT processes, such as o1-mini and o3-mini, outperform their non-long-CoT counterparts (e.g., gpt-4o-mini). For instance, on the Toys dataset, the NDCG@5 metric improves from 38.87 (for gpt-4o-mini) to 39.25 (for o1-mini) and 42.23 (for o3-mini). Similarly, gemini-2.0-flash-thinking-exp shows an improvement over the non-long-CoT gemini-2.0-flash, confirming that long-CoT reasoning provides a significant advantage in feature generation. While DeepSeek-V3 performs slightly worse than DeepSeek-R1 on the Instruments dataset, the latter outperforms DeepSeek-V3 on the larger Toys dataset.
   One notable exception is QwQ, which underperforms relative to the Qwen2.5-Instruct series. We speculate that this is due to QwQ generating fewer features, a point we will further explore in Section~\ref{sec:RQ3}.

\textbf{(3) Transfer of Advanced Reasoning to Recommendation:}
Models like DeepSeek-R1, gemini-2.0-flash-thinking-exp, and o3-mini stand out for their superior performance, with the first two excelling on different datasets. These models also perform strongly in math and coding topics of Chatbot Arena~\citep{chiang2024chatbot}, indicating that \textbf{the advanced inference scaling techniques driving their success in reasoning-intensive tasks also benefit recommendation}. 
We propose that this stems from the out-of-distribution (OOD) nature of recommendation data: due to highly personalized user behavior, recommendation data is generally not used during the pre-training of most LLMs~\citep{dubey2024llama, deepseek_v3}. As suggested by~\citep{yao2025unveiling}, training LLMs on long-CoT data can improve generalization to OOD tasks, enabling them to reason more effectively about nuanced user preferences.



In summary, we demonstrate that augmenting recommendation systems with LLM-generated features significantly improves performance, especially when models use long-CoT reasoning. Moreover, policy models that excel in math and coding tasks tend to generate superior features for capturing user preferences. This convergence underscores the potential of inference scaling as a general strategy for enhancing recommendation systems.

\subsection{Effect of Different Search Strategies}
\label{sec: RQ1_search}

In math and coding tasks, models often cannot arrive at a final solution in a single step. Instead, they often rely on search algorithms—either exploring multiple solution paths in parallel~\citep{brown2024large} or iteratively refining a candidate solution~\citep{madaan2024self}. Naturally, one might ask whether these same algorithms are equally beneficial for feature generation in a recommendation context, where the model's goal is to uncover user preferences. In this study, we evaluate four search strategies for generating features: CoT, Best-of-N, Beam Search, and MCTS. Details of each strategy can be found in Appendix~\ref{app:search_strategy}.

To balance effectiveness and computational efficiency, we evaluate these strategies on the smaller Instruments dataset, using Qwen2.5-7B-Instruct and Qwen2.5-14B-Instruct as our policy models. The reward model measures how many valid features—those that effectively distinguish between a user's liked and disliked items—each strategy generates, and we select the output with the highest count of valid features.

The results, presented in Table~\ref{tab:search_res}, show that all search strategies outperform the baseline CoT approach. For example, with Qwen2.5-14B-Instruct as the policy model, CoT alone achieves an NDCG@5 of 39.28. In comparison, advanced search algorithms achieve significantly better results: Best-of-N (43.11), Beam Search (41.09), and MCTS (40.64). When Qwen2.5-7B-Instruct is used in the ICL task, Beam Search slightly outperforms Best-of-N. However, as the policy model size increases (from 7B to 14B), Best-of-N maintains its advantage over step-level methods like Beam Search and MCTS. Overall, Best-of-N shows the largest performance gains especially when the policy model is sufficiently large, likely due to its ability to select the best features from multiple generated outputs.

These findings suggest that while step-level strategies like Beam Search and MCTS are effective in domains where precise incremental steps are crucial (such as solving mathematical proofs~\citep{xin2024deepseek} or playing games like Go~\citep{silver2016mastering}), Best-of-N is more effective in identifying the most relevant features for recommendation tasks. We will explore the reasons for this trend in greater detail in Section~\ref{sec: RQ3_search}.

\section{Advantages of Inference Scaling for Feature Augmentation(RQ2)}
\label{sec: RQ2}

In the previous section, we showed that features generated through inference scaling can significantly improve recommendation performance. However, it remains unclear how these gains are achieved—specifically, what advantages long-CoT LLMs offer over non-long-CoT models. 

\subsection{Increased Number of Unique Features}

To investigate this aspect, we compare the features generated by gpt-4o-mini and o1-mini.
We first collect all valid features generated across all users, then measure both the total number of features and, after clustering and removing duplicates, the number of unique features. In particular, we use the bge-m3 model~\citep{chen2024m3} to embed the features and DBSCAN~\citep{ester1996density} to group them. Since users can exhibit diverse decision factors~\citep{he2024impact}, deeper and more personalized reasoning allows the LLM to uncover the unique preferences and decision-making factors of each user. As the model performs more in-depth analysis of these varied factors, it is likely to generate a greater number of distinct features, each reflecting different aspects of the user's preferences. Therefore, the number of unique features can serve as a valuable metric for assessing the extent to which the model has captured the complexity and individuality of user decision-making.
The results, presented in Appendix Table~\ref{tab:feature_quantity}, show that o1-mini generates more features overall and more unique features than gpt-4o-mini, indicating that extended CoT reasoning enables a more comprehensive exploration of user preferences.

To further analyze the relationship between feature quantity and recommendation performance, we use the larger Toys dataset—chosen for its size, which minimizes noise and yields robust insights. In this dataset, we plot the recommendation performance of features generated by different policy models against the number of unique valid features (using the logarithm, base 10, of the count for clarity). The results, shown in Figure~\ref{fig:performance_wrt_count}, indicate a positive correlation: as the number of unique valid features increases, recommendation performance improves. This finding suggests that the quantity of unique valid features can be considered a reasonable indicator of the depth of personalized reasoning, and can serve as a useful predictor of recommendation performance.

\subsection{More Detailed and Specific Descriptions}

Beyond generating more features, we also observe that o1-mini produces features with more detailed and specific descriptions. To illustrate this more clearly, we provide an example in Appendix~\ref{app: appendix_example}. 

To rigorously evaluate the specificity of the descriptions, we adopt an LLM-as-a-judge~\citep{zheng2024judging} approach. 
In this method, we present pairs of feature descriptions—one from gpt-4o-mini and one from o1-mini—to three judge models (gpt-4o-mini, o1-mini, and claude-3-5-sonnet~\citep{ckaude_3_5_sonnet}).
To mitigate position bias~\citep{shi2024judging}, each feature pair is evaluated twice, with the positions of the generated features swapped between the two rounds. We consider a feature description superior only when both evaluations are consistent; otherwise, the result is recorded as a tie. 

\begin{figure}[t]
  \includegraphics[width=\columnwidth]{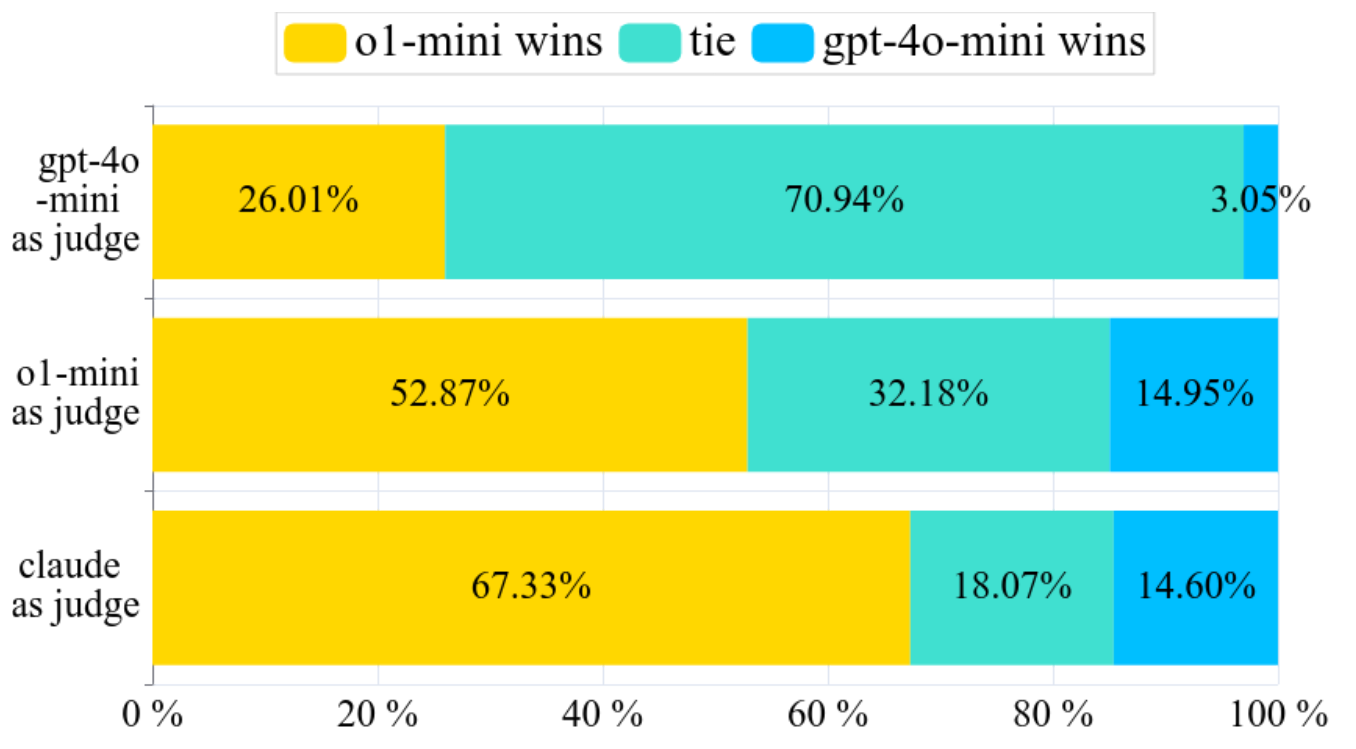}
  \caption{Win-Tie-Lose Comparisons on specificity of features from gpt-4o-mini and o1-mini.}
  \label{fig:llm_as_judge}
\end{figure}

The results are presented in Figure~\ref{fig:llm_as_judge}, indicating unanimous agreement among all judges that o1-mini generates more specific and better-described features than gpt-4o-mini. Notably, even though LLMs often exhibit self-enhancement bias~\citep{zheng2024judging, brown1986evaluations} — where they tend to favor their own generated content — gpt-4o-mini still rated o1-mini's descriptions more favorably. This further supports the conclusion that o1-mini produces more detailed and precise features.

In summary, the key benefits of inference scaling in feature generation are: (1) it generates more features, and (2) these generated features are more specific and detailed. These advantages suggest that longer reasoning chains enable models to explore a richer space of possible features, providing more meaningful insights into user preferences for recommendation tasks.

\section{Factors Influencing the Number of Generated Features (RQ3)}
\label{sec:RQ3}

Having established that inference scaling can produce both a larger and more concrete set of features, we now turn our attention to examining the factors that affect the quantity of generated features. 


\begin{figure*}[t]  
    \centering    

    \subfigure[]
    {
    \centering
    \includegraphics[width=0.48\linewidth]{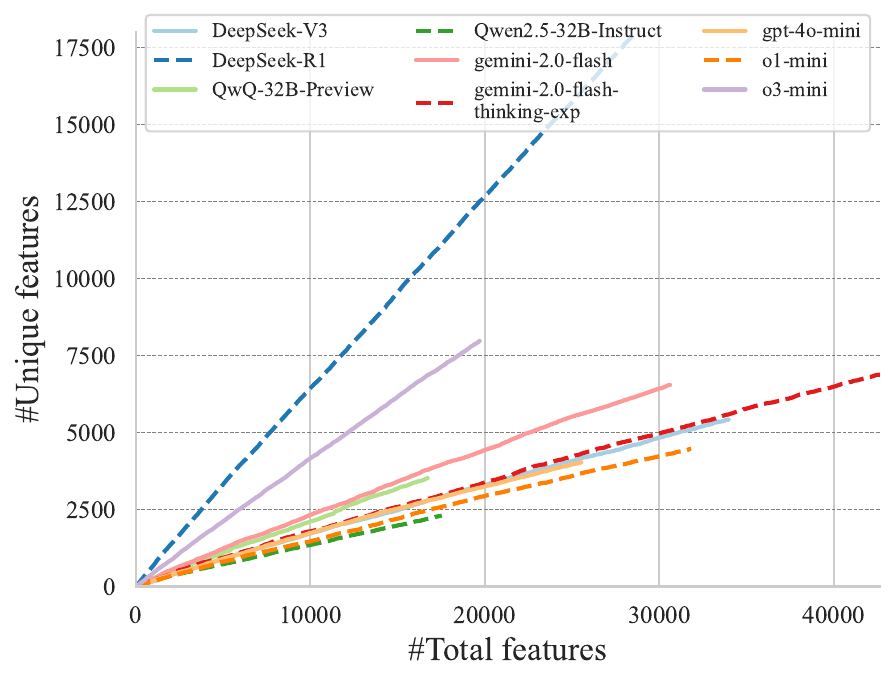}
    \label{fig: toy_multi_LLMs_a}
   }
  \subfigure[]
    {
    \centering
    \includegraphics[width=0.48\linewidth]{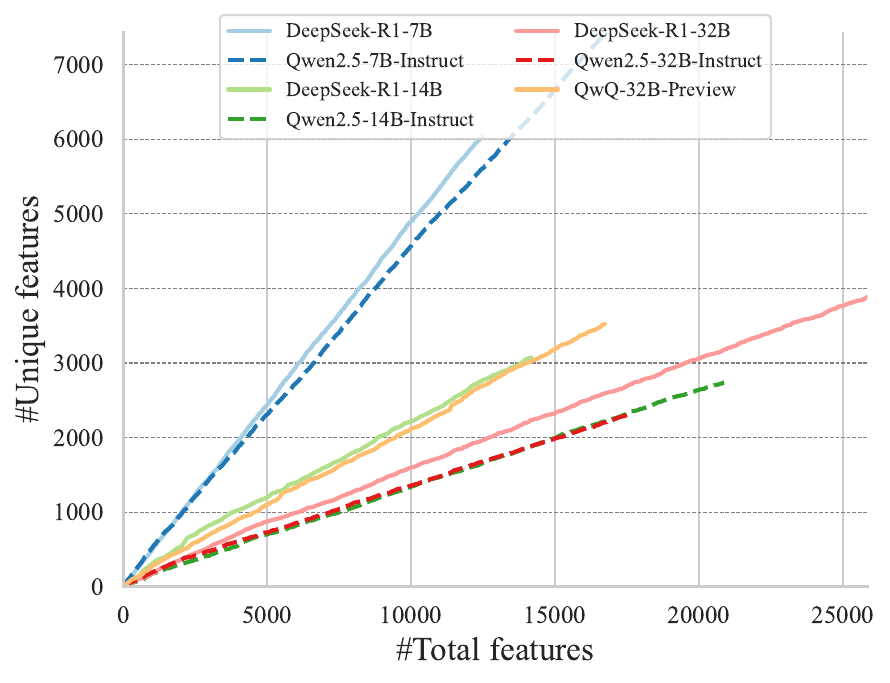}
    \label{fig: toy_multi_LLMs_b}
   }
   
    \caption{Number of unique features generated by different LLMs compared against the total number of generated features in the Toys dataset.}

    \label{fig:feature_growth_toy_multi_LLMs}
\end{figure*}

\subsection{Model Selection}

Following the procedure in Section~\ref{sec: RQ2}, we collect all features generated by each policy model across all users. The reward model then identifies which features effectively distinguish users’ liked items from disliked ones. We cluster and remove duplicates from these valid features, ultimately comparing the number of unique features from each policy model.

Figure~\ref{fig:feature_growth_toy_multi_LLMs} 
illustrates the growth of unique features as a function of the total number of valid features on the Toys dataset.
The horizontal axis represents the total number of valid, distinguishable features generated across all users, while the vertical axis indicates the number of unique features—serving as a proxy for the depth of personalization. Three key observations emerge:

\textbf{(1) Model Families.} The DeepSeek and Gemini families attain the top points on the vertical and horizontal axes, respectively. DeepSeek-R1 generates the highest number of unique valid features, whereas gemini-2.0-flash-thinking-exp produces a large number of features overall but fewer truly unique ones—indicating high quantity but somewhat weaker personalization. Within the GPT family, o3-mini stands out by producing a substantial number of unique features, second only to DeepSeek-R1. By contrast, the Qwen series yields fewer features in total (under 20K) and fewer unique features (under 5K). 

\textbf{(2) Long-CoT.} When comparing models within the same family, those incorporating extended reasoning—via SFT~\citep{o1_journey_part2}, RL~\citep{min2024imitate}, or other advanced training techniques~\citep{deepseek_r1}—produce significantly more unique features than their base counterparts. For example, in Figure~\autoref{fig: toy_multi_LLMs_a}, DeepSeek-R1 yields more than three times the number of unique features compared to DeepSeek-V3, while o3-mini outperforms gpt-4o-mini by over 50\%. Similarly, gemini-2.0-flash-thinking-exp surpasses gemini-2.0-flash, and the distilled versions of DeepSeek-R1 for Qwen exhibit steeper slopes than standard Qwen models of the same scale (Figure~\autoref{fig: toy_multi_LLMs_b}). 
We propose that explicitly producing a richer chain-of-thought—rather than keeping it implicit in model parameters—helps capture more granular decision factors, facilitating the discovery of a broader range of user-specific features. One exception is QwQ, which generates fewer features than Qwen2.5-7B-Instruct, aligning with the results in Section~\ref{sec: RQ1_policy} showing that QwQ underperforms Qwen in downstream recommendation tasks.

\textbf{(3) Model Size.} Overall, larger models tend to perform better, as exemplified by DeepSeek-R1 (with 671B parameters) generating over 17,500 unique features, whereas its distilled 7B, 14B, and 32B versions produce only a few thousand. Nonetheless, within the 7B–32B range, larger models do not always dominate: across both Qwen and distilled DeepSeek models, the 7B variants often surpass their 14B and 32B counterparts in personalization capacity. Intriguingly, smaller models that benefit from long-CoT training can even outperform larger models without extended reasoning. For instance, DeepSeek-V3 (671B) generates around 5K unique features, whereas DeepSeek-R1-7B produces over 7K—highlighting the substantial role of long chain-of-thought processes.


In summary, long-CoT training and model family exert the greatest influence on feature generation capacity. In highly personalized domains such as recommendation, where user decision factors vary widely, a detailed, case-by-case approach that leverages explicit and extended reasoning proves particularly advantageous.

\begin{figure}[t]  
    \centering    

    \subfigure[Qwen2.5-7B-Instruct as policy model]
    {
    \centering
    \includegraphics[width=0.45\linewidth]{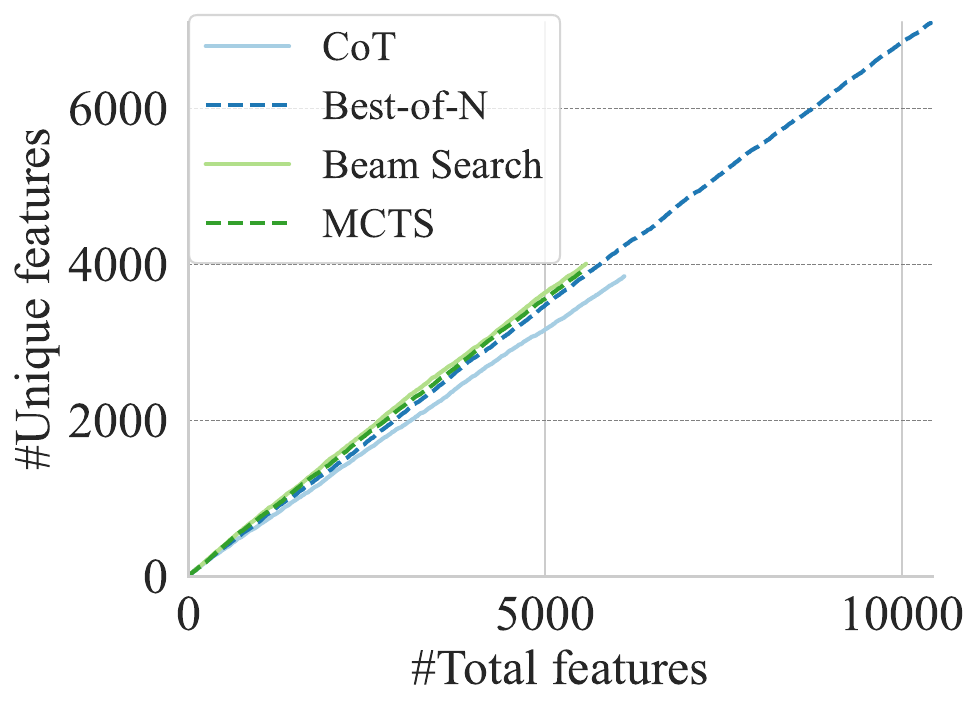}
    \label{fig: search_method_qwen7b}
   }
  \subfigure[Qwen2.5-14B-Instruct as policy model]
    {
    \centering
    \includegraphics[width=0.45\linewidth]{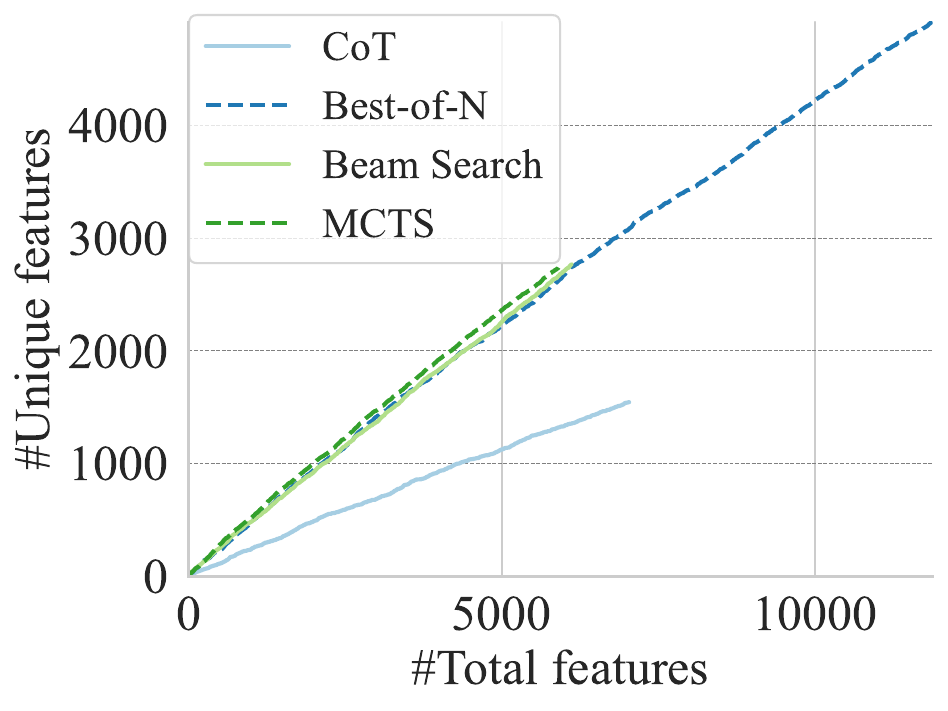}
    \label{fig: search_method_qwen14b}
   }
   
    \caption{Comparison of different search strategies on the Instruments dataset.}

    \label{fig:search_method}
\end{figure}

\subsection{Search Strategies}
\label{sec: RQ3_search}

We next investigate how different search strategies affect the quantity of generated features. Figure~\ref{fig:search_method} presents results on the Instruments dataset. All strategies exhibit a roughly linear relationship between the total number of generated features and the number of unique features. Notably, Best-of-N consistently produces the highest total feature count, while Beam Search and MCTS yield fewer features overall. This aligns with their recommendation performance in Section~\ref{sec: RQ1_search}, where Best-of-N outperforms the other search strategies.

The underlying reason lies in the difference between step-level and solution-level optimization. In tasks like math~\citep{xin2024deepseek} or game playing (e.g., Go~\citep{silver2016mastering}), each intermediate step must be correct to avoid disrupting the final outcome; local missteps can ruin the entire sequence, making step-level search essential for achieving a globally optimal result. In contrast, user-feature generation in recommendation is less tightly coupled: each newly proposed feature—whether relevant or not—does not necessarily affect subsequent features. Even if some features are ineffective, the model can still generate valid and valuable features afterward. For instance, generating features such as appearance, price, or durability are relatively independent, so discovering useful features after ineffective ones can still improve the overall result. Therefore, step-level search provides limited benefits in recommendation tasks, whereas solution-level strategies like Best-of-N are more effective in identifying the most promising solution without requiring each incremental step to be optimal.



\section{Related Work}


\textbf{LLMs for Rec.} LLMs can enhance recommendation systems by providing contextual understanding and reasoning abilities~\citep{huang2022towards}. Recent work on LLM-based augmentation in recommendation generally follows two main directions. The first approach leverages LLMs for text embedding, integrating user and item attributes into unified representations~\citep{hou2024bridging, zhang2024text, sheng2024language, lee2024star, yuan2023go}. The second approach focuses on generating additional information for recommendations. For instance, KAR~\citep{xi2024towards} instructs LLMs to produce item descriptions and user preference rationales, while LLM-ESR~\citep{liu2024llm} summarizes user preferences. RLMRec~\citep{ren2024representation} describes both user and item attributes, incorporating collaborative information, and LLM-CF~\citep{sun2024large} uses a chain-of-thought to represent user decisions. However, most of these methods rely on relatively fast inference, frequently resulting in incomplete coverage and insufficient specificity~\citep{ji2025test}.

\textbf{Inference Computation Scaling.} Inference scaling involves allocating more computational resources to the inference stage, allowing LLMs to reason based on additional prior steps. This shift from System-1 (fast thinking) to System-2 (slow thinking) reasoning enhances their ability to process complex tasks~\citep{ji2025test}. The relationship between performance and inference time was first examined by~\citet{snell2024Scaling}, and recent studies~\citep{qi2024mutual, zhang2024accessing, chen2024alphamath, yang2024qwen2, shao2024deepseekmath} have demonstrated promising results in math and coding tasks. Moreover,~\citet{yue2024inference} investigates inference scaling for Retrieval Augmented Generation, and~\citet{sun2025rearter} employs test-time scaling to address challenges in complex multi-step reasoning.
However, none of these works have investigated whether inference scaling can help address the incomplete coverage problem in recommendation tasks, and we aim to explore how this technique can improve personalized recommendation systems through feature augmentation.

\section{Conclusion}
In this paper, we demonstrate that inference scaling can significantly enhance feature augmentation for recommendation systems. 
Our experiments show that long-CoT models, compared to traditional fast inference methods, generate more detailed and specific features that better capture user decision factors, resulting in markedly improved recommendation performance.
In contrast to faster inference methods, long-CoT reasoning overcomes challenges like incomplete feature coverage and insufficient specificity. 
Additionally, we observe that the quantity and specificity of features are closely linked to the accuracy and effectiveness of the recommendations.  
Overall, our findings highlight the potential of inference scaling to improve user preference modeling and personalization in recommendation systems, paving the way for future research on its broader application across personalized tasks.


\section*{Limitations}

In this work, we demonstrate that inference scaling effectively enhances feature generation for recommendation tasks: extended chain-of-thought reasoning yields more diverse, specific features, leading to significant performance improvements. Nevertheless, several limitations warrant attention. First, inference scaling requires additional computational resources—both in terms of model size and inference time~\citep{snell2024Scaling}—which restricts its applicability in low-latency or resource-constrained settings. Second, although our empirical findings highlight the benefits of long-CoT reasoning, we do not provide a theoretical explanation of why inference scaling is so effective in recommendation tasks or why strategies such as beam search and MCTS do not offer comparable improvements. Finally, our experiments are limited to public datasets, so validating inference-scaled models in industrial-scale, heterogeneous environments with more complex user behaviors may introduce further challenges and necessitate refinements. Despite these limitations, our work provides valuable insights into the potential of inference scaling for personalized recommendation systems, underscoring a promising direction for future research.

\section*{Ethics Statement}
\textbf{Data Privacy and Security:}
All datasets used in this study are publicly available and adhere to the usage agreements provided by their respective data sources. The user information in these datasets is fully anonymized to protect individual privacy, and no personally identifiable information is disclosed.

\textbf{Fairness Consideration and Social Impact:}
We carefully controlled the use of LLMs to prevent bias or discrimination. Specifically, we avoided prompts that could intentionally mislead LLMs into generating harmful content. The prompts used in our experiments are made publicly available and can be verified. To further ensure safety, we manually checked a representative sample of generated outputs and employed LLM-based evaluation to screen for potentially offensive or biased content.

\textbf{Feature Generation Approach:} One of the primary goals of this research is to generate more specific features to improve recommendation quality. The features generated through our prompts are objective, without appraising users or items as ``good'' or ``bad.'' This approach minimizes the risk of introducing undue influence or bias into the recommendation process.

\textbf{Dataset Impact:} The datasets used in this study, which include user interactions with Toys and Instruments, further reduce the likelihood of negative social impact. The only user-related data we utilize are historical interactions and 1–5 ratings, limiting the scope of information available to the LLM. Given these constraints, harmful or sensitive content is highly unlikely to arise.



\bibliography{ref}

\begin{thebibliography}{62}
\providecommand{\natexlab}[1]{#1}

\bibitem[{Acharya et~al.(2023)Acharya, Singh, and Onoe}]{acharya2023llm}
Arkadeep Acharya, Brijraj Singh, and Naoyuki Onoe. 2023.
\newblock Llm based generation of item-description for recommendation system.
\newblock In \emph{Proceedings of the 17th ACM Conference on Recommender Systems}, pages 1204--1207.

\bibitem[{Anthropic(2024)}]{ckaude_3_5_sonnet}
Anthropic. 2024.
\newblock \href {https://www.anthropic.com/news/claude-3-5-sonnet} {Claude 3.5 sonnet.}

\bibitem[{Brown et~al.(2024)Brown, Juravsky, Ehrlich, Clark, Le, R{\'e}, and Mirhoseini}]{brown2024large}
Bradley Brown, Jordan Juravsky, Ryan Ehrlich, Ronald Clark, Quoc~V Le, Christopher R{\'e}, and Azalia Mirhoseini. 2024.
\newblock Large language monkeys: Scaling inference compute with repeated sampling.
\newblock \emph{arXiv preprint arXiv:2407.21787}.

\bibitem[{Brown(1986)}]{brown1986evaluations}
Jonathon~D Brown. 1986.
\newblock Evaluations of self and others: Self-enhancement biases in social judgments.
\newblock \emph{Social cognition}, 4(4):353--376.

\bibitem[{Browne et~al.(2012)Browne, Powley, Whitehouse, Lucas, Cowling, Rohlfshagen, Tavener, Perez, Samothrakis, and Colton}]{browne2012survey}
Cameron~B Browne, Edward Powley, Daniel Whitehouse, Simon~M Lucas, Peter~I Cowling, Philipp Rohlfshagen, Stephen Tavener, Diego Perez, Spyridon Samothrakis, and Simon Colton. 2012.
\newblock A survey of monte carlo tree search methods.
\newblock \emph{IEEE Transactions on Computational Intelligence and AI in games}, 4(1):1--43.

\bibitem[{Chen et~al.(2024{\natexlab{a}})Chen, Liao, Li, and Fan}]{chen2024alphamath}
Guoxin Chen, Minpeng Liao, Chengxi Li, and Kai Fan. 2024{\natexlab{a}}.
\newblock Alphamath almost zero: process supervision without process.
\newblock \emph{arXiv preprint arXiv:2405.03553}.

\bibitem[{Chen et~al.(2024{\natexlab{b}})Chen, Xiao, Zhang, Luo, Lian, and Liu}]{chen2024m3}
Jianlyu Chen, Shitao Xiao, Peitian Zhang, Kun Luo, Defu Lian, and Zheng Liu. 2024{\natexlab{b}}.
\newblock M3-embedding: Multi-linguality, multi-functionality, multi-granularity text embeddings through self-knowledge distillation.
\newblock In \emph{Findings of the Association for Computational Linguistics ACL 2024}, pages 2318--2335.

\bibitem[{Chiang et~al.(2024)Chiang, Zheng, Sheng, Angelopoulos, Li, Li, Zhang, Zhu, Jordan, Gonzalez et~al.}]{chiang2024chatbot}
Wei-Lin Chiang, Lianmin Zheng, Ying Sheng, Anastasios~Nikolas Angelopoulos, Tianle Li, Dacheng Li, Hao Zhang, Banghua Zhu, Michael Jordan, Joseph~E Gonzalez, et~al. 2024.
\newblock Chatbot arena: An open platform for evaluating llms by human preference.
\newblock \emph{arXiv preprint arXiv:2403.04132}.

\bibitem[{Cobbe et~al.(2021)Cobbe, Kosaraju, Bavarian, Chen, Jun, Kaiser, Plappert, Tworek, Hilton, Nakano et~al.}]{cobbe2021training}
Karl Cobbe, Vineet Kosaraju, Mohammad Bavarian, Mark Chen, Heewoo Jun, Lukasz Kaiser, Matthias Plappert, Jerry Tworek, Jacob Hilton, Reiichiro Nakano, et~al. 2021.
\newblock Training verifiers to solve math word problems.
\newblock \emph{arXiv preprint arXiv:2110.14168}.

\bibitem[{Dai et~al.(2023)Dai, Shao, Zhao, Yu, Si, Xu, Sun, Zhang, and Xu}]{dai2023uncovering}
Sunhao Dai, Ninglu Shao, Haiyuan Zhao, Weijie Yu, Zihua Si, Chen Xu, Zhongxiang Sun, Xiao Zhang, and Jun Xu. 2023.
\newblock Uncovering chatgpt’s capabilities in recommender systems.
\newblock In \emph{Proceedings of the 17th ACM Conference on Recommender Systems}, pages 1126--1132.

\bibitem[{Deepmind(2024)}]{gemini_2.0_flash}
Google Deepmind. 2024.
\newblock \href {https://blog.google/technology/google-deepmind/google-gemini-ai-update-december-2024/} {Introducing gemini 2.0: our new ai model for the agentic era}.

\bibitem[{Dubey et~al.(2024)Dubey, Jauhri, Pandey, Kadian, Al-Dahle, Letman, Mathur, Schelten, Yang, Fan et~al.}]{dubey2024llama}
Abhimanyu Dubey, Abhinav Jauhri, Abhinav Pandey, Abhishek Kadian, Ahmad Al-Dahle, Aiesha Letman, Akhil Mathur, Alan Schelten, Amy Yang, Angela Fan, et~al. 2024.
\newblock The llama 3 herd of models.
\newblock \emph{arXiv preprint arXiv:2407.21783}.

\bibitem[{Ester et~al.(1996)Ester, Kriegel, Sander, Xu et~al.}]{ester1996density}
Martin Ester, Hans-Peter Kriegel, J{\"o}rg Sander, Xiaowei Xu, et~al. 1996.
\newblock A density-based algorithm for discovering clusters in large spatial databases with noise.
\newblock In \emph{Proceedings of the Second International Conference on Knowledge Discovery and Data Mining}, page 226–231.

\bibitem[{Fang et~al.(2020)Fang, Zhang, Shu, and Guo}]{fang2020deep}
Hui Fang, Danning Zhang, Yiheng Shu, and Guibing Guo. 2020.
\newblock Deep learning for sequential recommendation: Algorithms, influential factors, and evaluations.
\newblock \emph{ACM Transactions on Information Systems (TOIS)}, pages 1--42.

\bibitem[{Geng et~al.(2022)Geng, Liu, Fu, Ge, and Zhang}]{geng2022recommendation}
Shijie Geng, Shuchang Liu, Zuohui Fu, Yingqiang Ge, and Yongfeng Zhang. 2022.
\newblock Recommendation as language processing (rlp): A unified pretrain, personalized prompt \& predict paradigm (p5).
\newblock In \emph{Proceedings of the 16th ACM Conference on Recommender Systems}, pages 299--315.

\bibitem[{Guo et~al.(2025)Guo, Yang, Zhang, Song, Zhang, Xu, Zhu, Ma, Wang, Bi et~al.}]{deepseek_r1}
Daya Guo, Dejian Yang, Haowei Zhang, Junxiao Song, Ruoyu Zhang, Runxin Xu, Qihao Zhu, Shirong Ma, Peiyi Wang, Xiao Bi, et~al. 2025.
\newblock Deepseek-r1: Incentivizing reasoning capability in llms via reinforcement learning.
\newblock \emph{arXiv preprint arXiv:2501.12948}.

\bibitem[{He et~al.(2017)He, Liao, Zhang, Nie, Hu, and Chua}]{he2017neural}
Xiangnan He, Lizi Liao, Hanwang Zhang, Liqiang Nie, Xia Hu, and Tat-Seng Chua. 2017.
\newblock Neural collaborative filtering.
\newblock In \emph{Proceedings of the 26th international conference on world wide web}, pages 173--182.

\bibitem[{He et~al.(2024)He, Liu, and Jung}]{he2024impact}
Xinyue He, Qi~Liu, and Sunho Jung. 2024.
\newblock The impact of recommendation system on user satisfaction: A moderated mediation approach.
\newblock \emph{Journal of Theoretical and Applied Electronic Commerce Research}, 19(1):448--466.

\bibitem[{Hou et~al.(2024)Hou, Li, He, Yan, Chen, and McAuley}]{hou2024bridging}
Yupeng Hou, Jiacheng Li, Zhankui He, An~Yan, Xiusi Chen, and Julian McAuley. 2024.
\newblock Bridging language and items for retrieval and recommendation.
\newblock \emph{arXiv preprint arXiv:2403.03952}.

\bibitem[{Huang and Chang(2022)}]{huang2022towards}
Jie Huang and Kevin Chen-Chuan Chang. 2022.
\newblock Towards reasoning in large language models: A survey.
\newblock \emph{arXiv preprint arXiv:2212.10403}.

\bibitem[{Huang et~al.(2024)Huang, Zou, Li, Liu, Zheng, Chern, Xia, Qin, Yuan, and Liu}]{o1_journey_part2}
Zhen Huang, Haoyang Zou, Xuefeng Li, Yixiu Liu, Yuxiang Zheng, Ethan Chern, Shijie Xia, Yiwei Qin, Weizhe Yuan, and Pengfei Liu. 2024.
\newblock O1 replication journey--part 2: Surpassing o1-preview through simple distillation, big progress or bitter lesson?
\newblock \emph{arXiv preprint arXiv:2411.16489}.

\bibitem[{Ji et~al.(2025)Ji, Li, Ye, Wu, Xu, Mo, and Zhang}]{ji2025test}
Yixin Ji, Juntao Li, Hai Ye, Kaixin Wu, Jia Xu, Linjian Mo, and Min Zhang. 2025.
\newblock Test-time computing: from system-1 thinking to system-2 thinking.
\newblock \emph{arXiv preprint arXiv:2501.02497}.

\bibitem[{Kahneman(2011)}]{kahneman2011thinking}
Daniel Kahneman. 2011.
\newblock Thinking, fast and slow.
\newblock \emph{Farrar, Straus and Giroux}.

\bibitem[{Lee et~al.(2024)Lee, Kraft, Jin, Mehta, Xu, Hong, Chi, and Yi}]{lee2024star}
Dong-Ho Lee, Adam Kraft, Long Jin, Nikhil Mehta, Taibai Xu, Lichan Hong, Ed~H Chi, and Xinyang Yi. 2024.
\newblock Star: A simple training-free approach for recommendations using large language models.
\newblock \emph{arXiv preprint arXiv:2410.16458}.

\bibitem[{Liu et~al.(2024{\natexlab{a}})Liu, Feng, Xue, Wang, Wu, Lu, Zhao, Deng, Zhang, Ruan et~al.}]{deepseek_v3}
Aixin Liu, Bei Feng, Bing Xue, Bingxuan Wang, Bochao Wu, Chengda Lu, Chenggang Zhao, Chengqi Deng, Chenyu Zhang, Chong Ruan, et~al. 2024{\natexlab{a}}.
\newblock Deepseek-v3 technical report.
\newblock \emph{arXiv preprint arXiv:2412.19437}.

\bibitem[{Liu et~al.(2023)Liu, Liu, Zhou, Lv, Zhou, and Zhang}]{liu2023chatgpt}
Junling Liu, Chao Liu, Peilin Zhou, Renjie Lv, Kang Zhou, and Yan Zhang. 2023.
\newblock Is chatgpt a good recommender? a preliminary study.
\newblock \emph{arXiv preprint arXiv:2304.10149}.

\bibitem[{Liu et~al.(2024{\natexlab{b}})Liu, Wu, Wang, Zhang, Tian, Zheng, and Zhao}]{liu2024llm}
Qidong Liu, Xian Wu, Yejing Wang, Zijian Zhang, Feng Tian, Yefeng Zheng, and Xiangyu Zhao. 2024{\natexlab{b}}.
\newblock Llm-esr: Large language models enhancement for long-tailed sequential recommendation.
\newblock In \emph{The Thirty-eighth Annual Conference on Neural Information Processing Systems}.

\bibitem[{Madaan et~al.(2024)Madaan, Tandon, Gupta, Hallinan, Gao, Wiegreffe, Alon, Dziri, Prabhumoye, Yang et~al.}]{madaan2024self}
Aman Madaan, Niket Tandon, Prakhar Gupta, Skyler Hallinan, Luyu Gao, Sarah Wiegreffe, Uri Alon, Nouha Dziri, Shrimai Prabhumoye, Yiming Yang, et~al. 2024.
\newblock Self-refine: Iterative refinement with self-feedback.
\newblock \emph{Advances in Neural Information Processing Systems}, 36.

\bibitem[{Min et~al.(2024)Min, Chen, Jiang, Chen, Deng, Hu, Tang, Wang, Cheng, Song et~al.}]{min2024imitate}
Yingqian Min, Zhipeng Chen, Jinhao Jiang, Jie Chen, Jia Deng, Yiwen Hu, Yiru Tang, Jiapeng Wang, Xiaoxue Cheng, Huatong Song, et~al. 2024.
\newblock Imitate, explore, and self-improve: A reproduction report on slow-thinking reasoning systems.
\newblock \emph{arXiv preprint arXiv:2412.09413}.

\bibitem[{Ni et~al.(2019)Ni, Li, and McAuley}]{ni2019justifying}
Jianmo Ni, Jiacheng Li, and Julian McAuley. 2019.
\newblock Justifying recommendations using distantly-labeled reviews and fine-grained aspects.
\newblock In \emph{Proceedings of the 2019 conference on empirical methods in natural language processing and the 9th international joint conference on natural language processing (EMNLP-IJCNLP)}, pages 188--197.

\bibitem[{OpenAI(2024{\natexlab{a}})}]{gpt_4o_mini}
OpenAI. 2024{\natexlab{a}}.
\newblock \href {https://openai.com/index/gpt-4o-mini-advancing-cost-efficient-intelligence/} {Gpt-4o mini: advancing cost-efficient intelligence}.

\bibitem[{OpenAI(2024{\natexlab{b}})}]{o1_blog}
OpenAI. 2024{\natexlab{b}}.
\newblock \href {https://openai.com/index/learning-to-reason-with-llms/} {Learning to reason with llms}.

\bibitem[{OpenAI(2024{\natexlab{c}})}]{o1_system_card}
OpenAI. 2024{\natexlab{c}}.
\newblock \href {https://cdn.openai.com/o1-system-card-20241205.pdf} {Openai o1 system card}.

\bibitem[{OpenAI(2025)}]{o3}
OpenAI. 2025.
\newblock \href {https://openai.com/index/openai-o3-mini/} {Openai o3-mini}.

\bibitem[{Qi et~al.(2024)Qi, Ma, Xu, Zhang, Yang, and Yang}]{qi2024mutual}
Zhenting Qi, Mingyuan Ma, Jiahang Xu, Li~Lyna Zhang, Fan Yang, and Mao Yang. 2024.
\newblock Mutual reasoning makes smaller llms stronger problem-solvers.
\newblock \emph{arXiv preprint arXiv:2408.06195}.

\bibitem[{Ren et~al.(2024)Ren, Wei, Xia, Su, Cheng, Wang, Yin, and Huang}]{ren2024representation}
Xubin Ren, Wei Wei, Lianghao Xia, Lixin Su, Suqi Cheng, Junfeng Wang, Dawei Yin, and Chao Huang. 2024.
\newblock Representation learning with large language models for recommendation.
\newblock In \emph{Proceedings of the ACM on Web Conference 2024}, pages 3464--3475.

\bibitem[{Shao et~al.(2024)Shao, Wang, Zhu, Xu, Song, Bi, Zhang, Zhang, Li, Wu et~al.}]{shao2024deepseekmath}
Zhihong Shao, Peiyi Wang, Qihao Zhu, Runxin Xu, Junxiao Song, Xiao Bi, Haowei Zhang, Mingchuan Zhang, YK~Li, Y~Wu, et~al. 2024.
\newblock Deepseekmath: Pushing the limits of mathematical reasoning in open language models.
\newblock \emph{arXiv preprint arXiv:2402.03300}.

\bibitem[{Sheng et~al.(2024)Sheng, Zhang, Zhang, Chen, Wang, and Chua}]{sheng2024language}
Leheng Sheng, An~Zhang, Yi~Zhang, Yuxin Chen, Xiang Wang, and Tat-Seng Chua. 2024.
\newblock Language representations can be what recommenders need: Findings and potentials.
\newblock \emph{arXiv preprint arXiv:2407.05441}.

\bibitem[{Shi et~al.(2024)Shi, Ma, Liang, Ma, and Vosoughi}]{shi2024judging}
Lin Shi, Chiyu Ma, Wenhua Liang, Weicheng Ma, and Soroush Vosoughi. 2024.
\newblock Judging the judges: A systematic investigation of position bias in pairwise comparative assessments by llms.
\newblock \emph{arXiv preprint arXiv:2406.07791}.

\bibitem[{Silver et~al.(2016)Silver, Huang, Maddison, Guez, Sifre, Van Den~Driessche, Schrittwieser, Antonoglou, Panneershelvam, Lanctot et~al.}]{silver2016mastering}
David Silver, Aja Huang, Chris~J Maddison, Arthur Guez, Laurent Sifre, George Van Den~Driessche, Julian Schrittwieser, Ioannis Antonoglou, Veda Panneershelvam, Marc Lanctot, et~al. 2016.
\newblock Mastering the game of go with deep neural networks and tree search.
\newblock \emph{nature}, 529(7587):484--489.

\bibitem[{Snell et~al.(2024)Snell, Lee, Xu, and Kumar}]{snell2024Scaling}
Charlie Snell, Jaehoon Lee, Kelvin Xu, and Aviral Kumar. 2024.
\newblock Scaling llm test-time compute optimally can be more effective than scaling model parameters.
\newblock \emph{arXiv preprint arXiv:2408.03314}.

\bibitem[{Sun et~al.(2024)Sun, Si, Zang, Zheng, Song, Zhang, and Xu}]{sun2024large}
Zhongxiang Sun, Zihua Si, Xiaoxue Zang, Kai Zheng, Yang Song, Xiao Zhang, and Jun Xu. 2024.
\newblock Large language models enhanced collaborative filtering.
\newblock In \emph{Proceedings of the 33rd ACM International Conference on Information and Knowledge Management}, pages 2178--2188.

\bibitem[{Sun et~al.(2025)Sun, Wang, Yu, Zang, Zheng, Xu, Zhang, Yang, and Li}]{sun2025rearter}
Zhongxiang Sun, Qipeng Wang, Weijie Yu, Xiaoxue Zang, Kai Zheng, Jun Xu, Xiao Zhang, Song Yang, and Han Li. 2025.
\newblock Rearter: Retrieval-augmented reasoning with trustworthy process rewarding.
\newblock \emph{arXiv preprint arXiv:2501.07861}.

\bibitem[{Wang et~al.(2024)Wang, Fang, Wan, Wen, Zhu, Liu, Gong, Song, Chen, Ni et~al.}]{wang2024openr}
Jun Wang, Meng Fang, Ziyu Wan, Muning Wen, Jiachen Zhu, Anjie Liu, Ziqin Gong, Yan Song, Lei Chen, Lionel~M Ni, et~al. 2024.
\newblock Openr: An open source framework for advanced reasoning with large language models.
\newblock \emph{arXiv preprint arXiv:2410.09671}.

\bibitem[{Wang and Lim(2023)}]{wang2023zero}
Lei Wang and Ee-Peng Lim. 2023.
\newblock Zero-shot next-item recommendation using large pretrained language models.
\newblock \emph{arXiv preprint arXiv:2304.03153}.

\bibitem[{Wang et~al.(2023)Wang, Chu, Ouyang, Wang, Hao, Shen, Gu, Xue, Zhang, Cui et~al.}]{wang2023enhancing}
Yan Wang, Zhixuan Chu, Xin Ouyang, Simeng Wang, Hongyan Hao, Yue Shen, Jinjie Gu, Siqiao Xue, James~Y Zhang, Qing Cui, et~al. 2023.
\newblock Enhancing recommender systems with large language model reasoning graphs.
\newblock \emph{arXiv preprint arXiv:2308.10835}.

\bibitem[{Wei et~al.(2022)Wei, Wang, Schuurmans, Bosma, Xia, Chi, Le, Zhou et~al.}]{wei2022chain}
Jason Wei, Xuezhi Wang, Dale Schuurmans, Maarten Bosma, Fei Xia, Ed~Chi, Quoc~V Le, Denny Zhou, et~al. 2022.
\newblock Chain-of-thought prompting elicits reasoning in large language models.
\newblock \emph{Advances in neural information processing systems}, 35:24824--24837.

\bibitem[{Wei et~al.(2024)Wei, Ren, Tang, Wang, Su, Cheng, Wang, Yin, and Huang}]{wei2024llmrec}
Wei Wei, Xubin Ren, Jiabin Tang, Qinyong Wang, Lixin Su, Suqi Cheng, Junfeng Wang, Dawei Yin, and Chao Huang. 2024.
\newblock Llmrec: Large language models with graph augmentation for recommendation.
\newblock In \emph{Proceedings of the 17th ACM International Conference on Web Search and Data Mining}, pages 806--815.

\bibitem[{Wu et~al.(2024)Wu, Zheng, Qiu, Wang, Gu, Shen, Qin, Zhu, Zhu, Liu et~al.}]{wu2024survey}
Likang Wu, Zhi Zheng, Zhaopeng Qiu, Hao Wang, Hongchao Gu, Tingjia Shen, Chuan Qin, Chen Zhu, Hengshu Zhu, Qi~Liu, et~al. 2024.
\newblock A survey on large language models for recommendation.
\newblock \emph{World Wide Web}, 27(5):60.

\bibitem[{Xi et~al.(2024)Xi, Liu, Lin, Cai, Zhu, Zhu, Chen, Tang, Zhang, and Yu}]{xi2024towards}
Yunjia Xi, Weiwen Liu, Jianghao Lin, Xiaoling Cai, Hong Zhu, Jieming Zhu, Bo~Chen, Ruiming Tang, Weinan Zhang, and Yong Yu. 2024.
\newblock Towards open-world recommendation with knowledge augmentation from large language models.
\newblock In \emph{Proceedings of the 18th ACM Conference on Recommender Systems}, pages 12--22.

\bibitem[{Xie et~al.(2023)Xie, Kawaguchi, Zhao, Zhao, Kan, He, and Xie}]{xie2023self}
Yuxi Xie, Kenji Kawaguchi, Yiran Zhao, James~Xu Zhao, Min-Yen Kan, Junxian He, and Michael Xie. 2023.
\newblock Self-evaluation guided beam search for reasoning.
\newblock \emph{Advances in Neural Information Processing Systems}, 36:41618--41650.

\bibitem[{Xin et~al.(2024)Xin, Ren, Song, Shao, Zhao, Wang, Liu, Zhang, Lu, Du et~al.}]{xin2024deepseek}
Huajian Xin, ZZ~Ren, Junxiao Song, Zhihong Shao, Wanjia Zhao, Haocheng Wang, Bo~Liu, Liyue Zhang, Xuan Lu, Qiushi Du, et~al. 2024.
\newblock Deepseek-prover-v1. 5: Harnessing proof assistant feedback for reinforcement learning and monte-carlo tree search.
\newblock \emph{arXiv preprint arXiv:2408.08152}.

\bibitem[{Yang et~al.(2024{\natexlab{a}})Yang, Yang, Zhang, Hui, Zheng, Yu, Li, Liu, Huang, Wei et~al.}]{qwen_2.5}
An~Yang, Baosong Yang, Beichen Zhang, Binyuan Hui, Bo~Zheng, Bowen Yu, Chengyuan Li, Dayiheng Liu, Fei Huang, Haoran Wei, et~al. 2024{\natexlab{a}}.
\newblock Qwen2.5 technical report.
\newblock \emph{arXiv preprint arXiv:2412.15115}.

\bibitem[{Yang et~al.(2024{\natexlab{b}})Yang, Zhang, Hui, Gao, Yu, Li, Liu, Tu, Zhou, Lin et~al.}]{yang2024qwen2}
An~Yang, Beichen Zhang, Binyuan Hui, Bofei Gao, Bowen Yu, Chengpeng Li, Dayiheng Liu, Jianhong Tu, Jingren Zhou, Junyang Lin, et~al. 2024{\natexlab{b}}.
\newblock Qwen2. 5-math technical report: Toward mathematical expert model via self-improvement.
\newblock \emph{arXiv preprint arXiv:2409.12122}.

\bibitem[{Yao et~al.(2025)Yao, Ren, Liao, and Liu}]{yao2025unveiling}
Xinhao Yao, Ruifeng Ren, Yun Liao, and Yong Liu. 2025.
\newblock Unveiling the mechanisms of explicit cot training: How chain-of-thought enhances reasoning generalization.
\newblock \emph{arXiv preprint arXiv:2502.04667}.

\bibitem[{Yuan et~al.(2023)Yuan, Yuan, Song, Li, Fu, Yang, Pan, and Ni}]{yuan2023go}
Zheng Yuan, Fajie Yuan, Yu~Song, Youhua Li, Junchen Fu, Fei Yang, Yunzhu Pan, and Yongxin Ni. 2023.
\newblock Where to go next for recommender systems? id-vs. modality-based recommender models revisited.
\newblock In \emph{Proceedings of the 46th International ACM SIGIR Conference on Research and Development in Information Retrieval}, pages 2639--2649.

\bibitem[{Yue et~al.(2024)Yue, Zhuang, Bai, Hui, Jagerman, Zeng, Qin, Wang, Wang, and Bendersky}]{yue2024inference}
Zhenrui Yue, Honglei Zhuang, Aijun Bai, Kai Hui, Rolf Jagerman, Hansi Zeng, Zhen Qin, Dong Wang, Xuanhui Wang, and Michael Bendersky. 2024.
\newblock Inference scaling for long-context retrieval augmented generation.
\newblock \emph{arXiv preprint arXiv:2410.04343}.

\bibitem[{Zeng et~al.(2024)Zeng, Cheng, Yin, Wang, Li, Zhou, Guo, Huang, and Qiu}]{zeng2024scaling}
Zhiyuan Zeng, Qinyuan Cheng, Zhangyue Yin, Bo~Wang, Shimin Li, Yunhua Zhou, Qipeng Guo, Xuanjing Huang, and Xipeng Qiu. 2024.
\newblock Scaling of search and learning: A roadmap to reproduce o1 from reinforcement learning perspective.
\newblock \emph{arXiv preprint arXiv:2412.14135}.

\bibitem[{Zhang et~al.(2024{\natexlab{a}})Zhang, Huang, Zhou, Li, and Ouyang}]{zhang2024accessing}
Di~Zhang, Xiaoshui Huang, Dongzhan Zhou, Yuqiang Li, and Wanli Ouyang. 2024{\natexlab{a}}.
\newblock Accessing gpt-4 level mathematical olympiad solutions via monte carlo tree self-refine with llama-3 8b.
\newblock \emph{arXiv preprint arXiv:2406.07394}.

\bibitem[{Zhang et~al.(2024{\natexlab{b}})Zhang, Bao, Yan, Wang, Feng, and He}]{zhang2024text}
Yang Zhang, Keqin Bao, Ming Yan, Wenjie Wang, Fuli Feng, and Xiangnan He. 2024{\natexlab{b}}.
\newblock Text-like encoding of collaborative information in large language models for recommendation.
\newblock \emph{arXiv preprint arXiv:2406.03210}.

\bibitem[{Zhang et~al.(2024{\natexlab{c}})Zhang, Wu, Yang, Shu, Xiao, Kong, and Sang}]{o1-coder}
Yuxiang Zhang, Shangxi Wu, Yuqi Yang, Jiangming Shu, Jinlin Xiao, Chao Kong, and Jitao Sang. 2024{\natexlab{c}}.
\newblock o1-coder: an o1 replication for coding.
\newblock \emph{arXiv preprint arXiv:2412.00154}.

\bibitem[{Zheng et~al.(2024)Zheng, Chiang, Sheng, Zhuang, Wu, Zhuang, Lin, Li, Li, Xing et~al.}]{zheng2024judging}
Lianmin Zheng, Wei-Lin Chiang, Ying Sheng, Siyuan Zhuang, Zhanghao Wu, Yonghao Zhuang, Zi~Lin, Zhuohan Li, Dacheng Li, Eric Xing, et~al. 2024.
\newblock Judging llm-as-a-judge with mt-bench and chatbot arena.
\newblock \emph{Advances in Neural Information Processing Systems}, 36.

\end{thebibliography}

\appendix

\section{Appendix}

\subsection{Search Strategies}
\label{app:search_strategy}

In this section, we introduce the search algorithms in detail:

\textbf{(a) CoT}. LLM explicitly outputs its intermediate reasoning steps. This has been the default approach in our experiments.

\textbf{(b) Best-of-N~\citep{cobbe2021training}.} The policy model generates $N$ complete outputs 
for each user. For each output, the reward model then evaluates each proposed feature inside and marks it as \emph{valid} if it effectively distinguishes between the user's liked and disliked items. Only the valid features are retained in each output, and the final output is selected as the one containing the greatest number of valid features.

\begin{table}
\centering
\caption{Statistics of the two datasets. }
\label{tab:stats_datasets}
\resizebox{0.8\linewidth}{!}{
\begin{tabular}{lccc}
\toprule
Dataset~                       & \#Users & \#Items & \#Interactions~  \\
\midrule
Toys    & 3,962  & 11,119  & 65,099          \\
Instruments & 1,411  & 6,317   & 23,804        \\
\bottomrule
\end{tabular}
}
\end{table}

\begin{table}
\centering
\caption{Total and unique feature counts generated by two policy models on different datasets.}
\label{tab:feature_quantity}
\resizebox{\linewidth}{!}{
\begin{tabular}{lccc}
\toprule
Dataset~        &        Policy Model~                       & \#Total features & \#Unique features   \\
\midrule
\multirow[m]{2}{*}{Toys} & gpt-4o-mini    & 25,536  & 4,037         \\
& o1-mini & 31,822   & 4,482        \\
\midrule
\multirow[m]{2}{*}{Instruments} & gpt-4o-mini    & 8,156  & 2,766          \\
& o1-mini & 11,443   & 4,086        \\
\bottomrule
\end{tabular}
}
\end{table}

\begin{lrbox}{0}
\begin{tcolorbox}[colback=green!5,colframe=green!35!black,boxrule=0.5pt,title={\textbf{LLM Prompt and Response Example}}]
\textbf{Prompt:}
{You are a user behavior analyst. Now a [user ID]  interacted with the following products:

[item 1], [item description 1], [user rating].

[item 2], [item description 2], [user rating].

…, …, …

What do you think is the biggest difference in attributes between these items that this user has rated high and low?

Return the most likely feature you can think of and the definition of this feature.
}\\

\textbf{Response:}
{Based on the historical user behavior provided, I think the reasons why users make these decisions can be attributed to:

\textbf{1. Durability Enhancements}: ….

\textbf{2. Brand Compatibility}: ….

…}
\end{tcolorbox}
\end{lrbox}

\begin{figure}[t]
    \centering
    \resizebox{0.9\linewidth}{!}{
        \usebox{0}
    }
    \caption{An example of a prompt for the policy model and its corresponding response.}
    \label{fig:PromptTemplate}
    \vspace{-15pt}
\end{figure}

\textbf{(c) Beam Search~\citep{xie2023self}.} The policy model first produces $N$ partial outputs (beams). The reward model then filters out the beams with lower rewards, retaining only $N/M$ of them. Each of these retained beams is then expanded by $M$ steps, maintaining a total of $N$ partial outputs at each stage. This process repeats until the final output is formed.

\textbf{(d) MCTS~\citep{browne2012survey}.} This lookahead method constructs a search tree from the root (initial prompt) to a terminal state (complete feature set). At each step, four phases occur: 
\begin{itemize}
    \item \textbf{Selection}: From the root, the algorithm traverses down to a leaf node that has not yet been explored; if all are explored, it chooses the leaf with the highest UCT (Upper Confidence Bound for Trees) score.
    \item \textbf{Expansion}: It expands the leaf node by generating one step forward.
    \item \textbf{Evaluation} (Rollout): From the newly added node, it simulates a path to a terminal state to estimate the reward via the reward model.
    \item \textbf{Backpropagation}: The reward is propagated back to update the intermediate nodes, guiding future selections.
\end{itemize} 

CoT and Best-of-N operate on the \textbf{solution level}, selecting among multiple fully generated outputs. In contrast, beam search and MCTS function at the \textbf{step level}, intervening on the policy model's partial outputs as they unfold~\citep{zeng2024scaling}.

\subsection{Example of specificity Comparison}
\label{app: appendix_example}
Here we provide a comparison of features generated by gpt-4o-mini and o1-mini in response to the same user request, as shown in Figure~\ref{fig:4o_o1_comparison_example}. As highlighted in red, gpt-4o-mini produces a relatively vague feature—``Component Functionality''—while o1-mini yields more specific attributes, such as ``Component Type'', ``surface'', ``material'', and ``performance or usability''. This example illustrates how extended CoT reasoning in o1-mini can lead to more detailed feature descriptions.


\onecolumn

\begin{lrbox}{1}
\begin{tcolorbox}[colback=green!5,colframe=green!35!black,boxrule=0.5pt,title={\textbf{Features generated by gpt-4o-mini}}]
{
"Instrument Quality": "Reflects the perceived quality and material of the instrument components based on user feedback. Possible values: `High', `Medium', `Low'",

"Usability": "Indicates how easy or difficult the instrument components are to use or install based on the user's experience. Possible values: `Easy', `Moderate', `Difficult'",

"Sound Characteristics": "Describes the tonal qualities and output levels of the instrument components, specifically for pickups. Possible values: `Bright', `Mellow', `Aggressive', `Balanced'",

"Brand Reputation": "The perceived trust and quality associated with specific brands based on user history. Possible values: `Highly Regarded', `Moderately Regarded', `Low Reputation'",

"User Preferences": "Denotes specific features that align with the user's musical style or performance preferences, like compatibility with certain styles. Possible values: `Suits Rock', `Suits Jazz', `Suits Metal', `Suits Blues'",

"Component Functionality": "Indicates the \textcolor{red}{functional aspects of components}, such as reliability and effectiveness in enhancing the overall performance of an instrument. Possible values: `Highly Functional', `Moderately Functional', `Low Functionality'"
}
\end{tcolorbox}
\end{lrbox}

\begin{lrbox}{2}
\begin{tcolorbox}[colback=green!5,colframe=green!35!black,boxrule=0.5pt,title={\textbf{Features generated by o1-mini}}]
{
"Component Type": "\textcolor{red}{The category of the instrument part}, indicating its primary function. Possible values include: Switch, Screws, Washer, Knobs, Pickups, Bridge Assembly, Strap Hardware, etc.",

"Aesthetic Finish": "The visual appearance and \textcolor{red}{surface treatment of the component}. Possible values include: Chrome, White, Black, Silver, Matte, Glossy, etc.",

"Material Composition": "\textcolor{red}{The materials} used in the construction of the component. Possible values include: Stainless Steel, Plastic, Felt, Metal Alloy, Aluminum, Brass, etc.",

"Functional Features": "Special features that enhance the \textcolor{red}{component's performance or usability}. Possible values include: Wax Potted, Heavy-Duty, Locking Mechanism, Adjustable, etc.",

"Sound Characteristics": "For components related to sound production, describes the tonal qualities. Possible values include: Aggressive, Clear, Loud, Obnoxious, Warm, Bright, etc.",

"Installation Complexity": "The ease or difficulty of installing the component. Possible values include: Easy, Moderate, Difficult, Requires Professional Installation, Tool-Free, etc.",

"Durability Enhancements": "Features that improve the longevity and resilience of the component. Possible values include: Corrosion-Resistant, Reinforced, Shock-Absorbing, Wear-Resistant, etc.",

"Brand Compatibility": "The compatibility of the component with specific brands or models. Possible values include: Fender Stratocaster, Gibson Les Paul, Universal Fit, Artist Signature Models, etc.",

"Customization Options": "Available options for personalizing the component. Possible values include: Number of Positions (e.g., 3-way, 5-way), Color Variants, Finish Types, etc.",

"Usage Context": "The intended use or suitable musical genres for the component. Possible values include: Rock, Metal, Jazz, Blues, Punk, Nu-Metal, etc."}
\end{tcolorbox}
\end{lrbox}

    


\begin{figure*}[htbp]
    \centering
    
    \subfigure[]{
        \centering
        \resizebox{0.9\textwidth}{!}{
            \usebox{1}
        }
        \label{fig:feature_4o_example}
    }
    \hfill
    \subfigure[]{
        \centering
        \resizebox{0.9\textwidth}{!}{
            \usebox{2}
        }
        \label{fig:feature_o1_example}
    }

    \caption{Comparison of features generated by gpt-4o-mini and o1-mini.}
    \label{fig:4o_o1_comparison_example}
\end{figure*}

\twocolumn

\end{document}